# Role of the Three-Body Interactions in the Ground-States of $^{3}$H and $^{4}$He Nuclei

S. B. Doma[1)] and H. S. El-Gendy[2)]

[1)]Faculty of Science, Alexandria University, Alexandria, Egypt, sbdoma@yahoo.com.
[2)]Faculty of Science & Art, Shaqra University, Shaqra, KSA, hselgendy1@yahoo.com.

**Abstract**
The role of the three-body interactions in the ground-states of $^{3}$H and $^{4}$He nuclei has been investigated by using two different methods. Accordingly, the ground-state nuclear wave functions, the binding energies, the root mean-square radii and the first excited-state energies of the $^{3}$H and $^{4}$He nuclei are investigated by applying the translation invariant shell model with basis functions corresponding to even number of quanta of excitations in the range $0 \leq N \leq 20$ and using two residual two-body interactions, given by the first author, together with three-body interaction in the form of a delta force. Furthermore, we have calculated the binding energy and the root mean-square radius of these nuclei by applying the variational Monte Carlo method and using the Reid $V_8$ two-body potential together with the Urbana model of the three-nucleon interaction.

## 1. Introduction

The nuclear shell model [1] has achieved wide popularity among nuclear theorists owing to its easily visualizable character and the success in interpreting experimental results. The assumption that, to a first approximation, each nucleon moves in an average potential independent of the motion of the other nucleons is an attractive one, due to one's familiarity with the Hartree fock theory of atomic structure.

The methods of expanding the nuclear wave function in terms of a complete set of orthonormal functions, basis functions, have been used on a large scale especially for the nuclei with $3 \leq A \leq 6$ [2].

The translation invariant shell model [2-5] (TISM) has shown good results for the structure of light nuclei with $A \leq 7$ by using nucleon-nucleon interactions [5-10]. This model considers the nucleus as a system of noninteracting quasi particles and enables us to apply the algebraic methods for studying the general features of matrix elements of operators that correspond to physical quantities. The TISM is based on the group theoretical methods of classifying the basis functions. The basis functions of this model are constructed in such a way that they will have certain symmetry with respect to the interchange of particles and have definite total angular momentum $J$ and isotopic spin $T$. The basis functions of the TISM are then expanded in the form of products of two types of functions, one corresponding to the set of A-2 nucleons and the other corresponding to the last pair of nucleons by means of the two-particle fractional parentage coefficients [2,6]. Using two-body interactions it is then possible to calculate the Hamiltonian matrices of the different nuclear states. In principle, the predicted results for the nuclear characteristics should be independent of the particular chosen basis functions when the number of terms in the expansion is kept large enough. The inclusion of all such bases in the expansion is too difficult since the matrices of the two-particle fractional parentage coefficients corresponding to these basis functions grow rapidly. It is therefore fundamental to have some rules that would allow us to reduce the number of these bases. Some of these rules are adopted for the nuleus $^6$Li in ref. [6].

On the other hand, microscopic calculations of light nuclei and nuclear matter [11] have indicated that it is difficult to explain the observed binding-energies and densities if we assume a non relativistic nuclear Hamiltonian having only two-nucleon interactions, consistent with the nucleon-nucleon scattering data at low energies ($E_{lab} < \sim 400 MeV$).

Since nucleons are composite objects made up of quarks and gluons, we can not approximate their interactions by a sum of two-body terms. The mesonic degrees of freedom can also generate three- and more-body potentials in the Hamiltonian in which only the nucleon degrees of freedom are retained. Since the energies obtained with Hamiltonian having only two-body potentials are not far from the experiment, we expect that the contribution of the many-body potentials is small compared to that of the two-body interaction in the realm of nuclear physics, and particularly, only three body-potentials may be important.

The effect of the three-nucleon interactions has been studied recently [12], where the effect of the different three-nucleon interactions in $p$-$^3$He elastic scattering at low energies has been calculated for the four-nucleon scattering observables by using the Kohn variational principle and the hyperspherical harmonics technique. On the other hand, the effects of the two-body and the three-body hyperon-nucleon interactions in $\Lambda$ hypernuclei have been stydied by assessing the relative importance of two- and three-body hyperon-nucleon force and by studying the effect of the hyperon-nucleon-nucleon interaction in closed shell $\Lambda$ hypernuclei from A = 5 to 91 [13,14]. Moreover, the authors in [15] extended the formalism of self-consistent Green's function theory to include three-body interactions and applied it to isotopic chains around oxygen for the first time. Furthermore, the authors in [16] used the realistic Argonne $v_{18}$ potential for the two-nucleon interaction and Urbana three-nucleon potentials to generate accurate variational Monte Carlo wave functions for the $A \leq 12$ nuclei.

The ab initio no-core shell model (NCSM) is a well-established theoretical framework aimed at an exact description of nuclear structure starting from high-precision interactions between the nucleons. Barrett, Navrátil, and Vary [17] discussed, in details, the extension of the ab initio NCSM to nuclear reactions and sketch a number of promising future directions for research emerging from the NCSM foundation, including a microscopic non-perturbative framework for the theory with a core. In the NCSM, C. Forssén and P. Navrátil [18] considered a system of A point-like, non-relativistic nucleons that interact by realistic inter-nucleon interactions. They considered two-nucleon interactions that reproduce nucleon-nucleon phase shifts with high precision, typically up to 350 MeV lap energy. Also, they included three-nucleon interactions with terms, e.g., related to two-pion exchanges with an intermediate delta excitation. Both semi-phenomenological potentials, based on meson-exchange models, as well as modern chiral interactions are considered.

In the present paper we have applied two different methods in order to investigate the role of the three-body interactions, together with the two-body interactions, in the ground-state characteristics of the $^3$H and $^4$He nuclei. For these nuclei we have investigated the ground-state energies and wave functions, the first excited state energies and the root mean-square radii.

In the first method we applied the TISM with basis functions corresponding to even number of quanta of excitations in the range $0 \leq N \leq 20$ and used two nucleon-nucleon interactions, given by the first author, which are constructed in the form of Gaussian interactions of the types given by Gogny, Pires and De Tourreil (GPT)- potential [19]. These potentials contain all the standard terms [20], namely, the central, the tensor, the spin-orbit, and the quadratic spin orbit terms, which are extremely suitable for the calculations of the different matrix elements with respect to the harmonic-oscillator basis functions. The parameters of these interactions are so chosen in such a way that they represent the long-range attraction and the short-range repulsion of the nucleon-nucleon interactions. These parameters are also chosen so as to reproduce good agreement between the calculated values of the binding energy, the root mean-square radius, the D-state probability, the magnetic dipole moment and the electric quadrupole moment of deuteron.

The methods of constructing the nuclear wave functions in the TISM and of calculating the matrix elements of the nucleon-nucleon interactions and the other one- and two-body operators are well explained in refs. [6,10]. Furthermore, we improved the obtained results by adding a three-body interaction of the form of Skyrme III potential [21] to the interaction Hamiltonian.

In the second method we have applied the variational Monte Carlo (VMC) method with suitable trial wave functions. In the calculations we have used two- and three-body interactions. For the two-body interactions we used the Reid $V_8$ model [22] by assuming that the interaction in all isospin-spin $T, S = 1, 0; 1, 1; 0, 0$ and $0, 1$ states is given by that in the $^1S_0$, ($^3P_2$ - $^3F_2$), $^1P_1$ and ($^3S_1$ - $^3D_1$) channels. This potential fits the scattering data in $^1S_0$, ($^3S_1$ - $^3D_1$), $^1P_1$ and ($^3P_2$ - $^3F_2$) channels and it does not have any quadratic spin orbit terms. For the three-body interaction we used the Urbana model three-nucleon potential (UVII), which has been written in the form of sum of a long-range two pion exchange and intermediate-range repulsive terms. The radial functions of this potential are associated with the tensor and Yukawa parts of one pion-exchange part [23].

The variational wave function is taken to be of the form $\left(S \prod_{i<j} F_{ij}\right)\emptyset$, where $S$ is a symmetrizing operator, the pair correlation operators $F_{ij}$ can include central, spin-orbit and tensor correlations, and $\emptyset$ is a pure spin-isospin function with no spatial dependence [23]. The general form of this wave function can be obtained by multiplying it by triplet-correlation operators, which include components induced by three-nucleon potentials.

**2. Calculations Using the Translation Invariant Shell Model**

The internal Hamiltonian of a nucleus consisting of $A$ nucleons, interacting via two-body potential, can be written in terms of the relative coordinates of the nucleons, in the form [8,10,24]

$$H = H^{(0)} + V', \qquad (2.1)$$

where

$$H^{(0)} = \frac{1}{A}\sum_{1=i<j}^{A}\left[\frac{(\boldsymbol{p}_i - \boldsymbol{p}_j)^2}{2m} + \frac{1}{2}m\omega^2(\boldsymbol{r}_i - \boldsymbol{r}_j)^2\right], \qquad (2.2)$$

is the well-known translation invariant shell model (TISM) Hamiltonian and

$$V' = \sum_{1=i<j}^{A}\left[V(|\boldsymbol{r}_i - \boldsymbol{r}_j|) - \frac{m\omega^2}{2A}(\boldsymbol{r}_i - \boldsymbol{r}_j)^2\right]. \qquad (2.3)$$

is the residual two-body interaction.

The energy eigenfunctions and eigenvalues of the Hamiltonian $H^{(0)}$ are given by [8,10]

$$|A\,\Gamma; M_L M_S T M_T\rangle \equiv |A\,N\{\rho\}(v)\alpha[f]LS; M_L M_S T M_T\rangle, \qquad (2.4)$$

$$E_N^{(0)} = \left\{N + \frac{3}{2}(A-1)\right\}\hbar\omega. \qquad (2.5)$$

In eq. (2.4) $\Gamma$ stands for the set of all irreducible group representations $N$, $\{\rho\}$, $(v)$, $\alpha$, $[f]$, $L$ and $S$. The functions (2.4) form a complete set of functions, bases. It is easy to construct bases which have definite total momentum $J$ in the form [8,10]

$$|A\,\Gamma J M_J T M_T\rangle = \sum_{M_L+M_S=M_J}(L M_L S M_S|J M_J)|A\,\Gamma; M_L M_S T M_T\rangle, \qquad (2.6)$$

where $(L M_L, S M_S|J M_J)$ are the Clebsch-Gordan coefficients of the rotational group $R_3$. The nuclear wave function of a state with total momentum $J$, isospin $T$ and parity $\pi$ can be constructed as follows [8,10]

$$|A J^\pi T M_J M_T\rangle = \sum_\Gamma C_\Gamma^{J^\pi T}|A\,\Gamma J M_J T M_T\rangle, \qquad (2.7)$$

where $C_\Gamma^{J^\pi T}$ are the state-expansion coefficients. In the summation (2.7) the number of quanta of excitations $N$ is permitted to be either even or odd integer depending on the parity of the state $\pi$.

The matrix elements of the residual interaction $V'$ with respect to the bases (2.6) are given in details in [8,10,24,25]. The ground-state nuclear wave function, which is obtained as a consequence of the diagonalization of the ground-state energy matrix, is used to calculate the root mean-square radius, $R$, from the well-known formula, given in [10,24,25]

The two-nucleon interaction $V(|\mathbf{r}_i - \mathbf{r}_j|)$, equation (2.3), which has been used in our calculations, has the well-known form [20]

$$V(r) = {}^{ts}X\{V_C(r) + V_T(r)S_{12} + V_{LS}\{r\}\boldsymbol{\ell}\cdot\mathbf{s} + V_{LL}(r)L_{12}\}, \qquad (2.8)$$

where $\boldsymbol{\ell}$, $\mathbf{s}$ and $t$ are the orbital angular momentum, the spin angular momentum and the isotopic spin of the two-nucleon state, respectively. The central, tensor, spin-orbit and quadratic spin-orbit terms are standard. The operator ${}^{ts}X$ has the form [20]

$$ {}^{ts}X = C_W + (-1)^{s+t+1}C_M + (-1)^{s+1}C_B + (-1)^{t+1}C_H, \qquad (2.9)$$

where $C_W$, $C_M$, $C_B$ and $C_H$ are the Wigner, the Majorana, the Bartlett and the Heisenberg exchange constants, respectively. Each term of the interaction is expressed as a sum of Gaussian functions in the form

$$V_\alpha(r) = \sum_{k=1}^{4} V_{\alpha k}\, e^{-\frac{r^2}{r_{\alpha k}^2}}, \qquad (2.10)$$

where $\alpha = C, T, LS$ and $LL$.

Two sets of values are considered for the exchange constants. For the first set we have: $C_W = 0.1333$, $C_M = -0.9333$, $C_B = -0.4667$ and $C_H = -0.2667$, which are known as the Rosenfeld constants and belong to the symmetric case. We refer to the potential resulting from this case by Pot-I. In the second set we have $C_W = -0.41$, $C_M = -0.41$, $C_B = -0.09$ and $C_H = 0.09$, which belong to the Serber case. The resulting potential is denoted by Pot-II. For the triplet-even state $(t = 0, s = 1)$, which is the case for the ground-state of deuteron, and from the normalization condition of the exchange constants the operator $^{ts}X$ equals -1, for both of the symmetric and the Serber cases so that the two types of the exchange constants will produce the same results for the ground-state characteristics of deuteron.

In the present paper we have varied the depth and range parameters $V_{\alpha k}$ and $r_{\alpha k}$, respectively, in order to obtain results for the binding energy, the root mean-square radius, the D-state probability, the magnetic dipole moment and the electric quadrupole moment of deuteron in excellent agreement with the corresponding experimental values.

It is well known that three-body forces are important to describe the properties of finite nuclei. The parameters in the nucleon-nucleon potential may not be unique or there may be some redundant parameters in order to reproduce the deuteron properties. In order to investigate these points of view, we have considered the following Hamiltonian operator, which takes into consideration the three-body forces:

$$H = H^{(0)} + V' + V'', \qquad (2.11)$$

where the first two terms in (2.11) are given by (2.2), and (2.3) and

$$V'' = \sum_{i,j,k} V(\boldsymbol{r_i}, \boldsymbol{r_j}, \boldsymbol{r_k}) \qquad (2.12)$$

is the three-body potential. For the three-body potential we have used the Skyrme-III potential [21]

$$V'' = t_3 \delta(\boldsymbol{r_1} - \boldsymbol{r_2})\delta(\boldsymbol{r_2} - \boldsymbol{r_3}), \qquad (2.13)$$

where $t_3 = 14000.0 \, MeV \times fm^6$.

To calculate the matrix elements of the three-body interactions, given by (2.13), we have used the fact that

$$\delta(\boldsymbol{r_1} - \boldsymbol{r_2}) = \frac{1}{r_1 r_2} \delta(r_1 - r_2)\delta(cos\theta_1 - cos\theta_2)\delta(\varphi_1 - \varphi_2) \qquad (2.14)$$

## 3. Calculations with the Variational Monte Carlo Method

In this method the nuclear Hamiltonian has the same form given by (2.11), $V'$ is a two-nucleon interaction that fits nucleon-nucleon (*NN*) scattering data and deuteron properties and $V''$ is an explicit three-body nucleon interaction. For the two-nucleon interactions we used the $V_8$–model potential [22]. This potential is obtained by assuming that the interaction in all isospin-spin: T, S =1, 0; 1, 1; 0, 0 and 0, 1 states is given by that in the $^1S_0$, ($^3P_2 - {^3F_2}$), $^1P_1$ and ($^3S_1 - {^3D_1}$) channels. This potential was introduced by Reid to fit the scattering data in $^1S_0$, ($^3S_1 - {^3D_1}$), $^1P_1$ and ($^3P_2 - {^3F_2}$) channels and is used to construct a two nucleon interaction operator

$$V_{8,ij} = V^C + V^\sigma(\boldsymbol{\sigma_i}.\boldsymbol{\sigma_j}) + V^\tau(\boldsymbol{\tau_i}.\boldsymbol{\tau_j}) + V^{\sigma\tau}(\boldsymbol{\sigma_i}.\boldsymbol{\sigma_j})(\boldsymbol{\tau_i}.\boldsymbol{\tau_j}) + V^t S_{ij} + V^{t\tau} S_{ij}(\boldsymbol{\tau_i}.\boldsymbol{\tau_j})$$

$$+V^b(\boldsymbol{L}.\boldsymbol{S_{ij}}) + V^{b\tau}(\boldsymbol{L}.\boldsymbol{S_{ij}}).(\boldsymbol{\tau_i}.\boldsymbol{\tau_j}) = \sum_{p=1,8} V^p(r_{ij})O^p_{ij}, \qquad (3.1)$$

where $V^p(r_{ij})$ are functions of the interparticle distance $r_{ij}$ and $O^p_{ij}$ are the operators

$$O^{p=1,8}_{ij} = 1, \boldsymbol{\sigma_i}.\boldsymbol{\sigma_j}, \boldsymbol{\tau_i}.\boldsymbol{\tau_j}, (\boldsymbol{\sigma_i}.\boldsymbol{\sigma_j})(\boldsymbol{\tau_i}.\boldsymbol{\tau_j}), S_{ij}, S_{ij}(\boldsymbol{\tau_i}.\boldsymbol{\tau_j}), (\boldsymbol{L}.\boldsymbol{S_{ij}}), (\boldsymbol{L}.\boldsymbol{S_{ij}})(\boldsymbol{\tau_i}.\boldsymbol{\tau_j}). \qquad (3.2)$$

Here $S_{ij}$ is the tensor operator and $L.S_{ij}$ is the spin-orbit operator.

Since this model does not have any quadratic spin-orbit terms the potentials in these channels were assumed to act in all S, T states 0, 1; 0, 0; 1, 0 and 1, 1 respectively. This model ignores differences between $^1S_0$ and $^1D_2$ potentials, and also assumes that the $^3P_0$ and $^3P_1$ potentials can be obtained from the potential operator in the $^3P_2$ - $^3F_2$ state.

The Reid potential has a modified one pion exchange part [26]

$$v_\pi = 3.488 \frac{e^{-x}}{x}\left(O^{\sigma\tau} + \left[\left(1 + \frac{3}{x} + \frac{3}{x^2}\right) - \left(\frac{12}{x} + \frac{3}{x^2}\right)e^{-3x}\right]O^{t\tau}\right), x = 0.7r. \quad (3.3)$$

The three-nucleon interaction used here is the Urbana-model three nucleon interaction given in Eqs. (4.1) and (4.2) of section-4.

**3.1 The Variational Monte Carlo Method**
The variational Monte Carlo (VMC) techniques have proven to be a very powerful tool for studying quantum systems, particularly within atomic [27-29] and nuclear physics [30]. The most important applications in nuclear physics are to few-body systems, either at the nucleon or the quark level. A variety of nuclear properties have been calculated with Monte Carlo methods, including the binding energies, the expectation values of other operators (besides $\widehat{H}$), electromagnetic form factors, and asymptotic properties of the wave function for three- and four-body nuclei. One can also study the Hamiltonian itself, in particular the effects of the three-nucleon interaction in light nuclei. Also in this technique we employ random sampling.

The variational Monte Carlo (VMC) method is used to evaluate expressions like [31]:

$$E_T = \frac{\int \Psi_T^*(R)\widehat{H}\,\Psi_T(R)dR}{\int \Psi_T^*(R)\Psi_T(R)dR} = \frac{\int \Psi_T^*(R)\,\Psi_T(R)dR[\Psi_T^{-1}(R)\widehat{H}\,\Psi_T(R)]}{\int \Psi_T^*(R)\,\Psi_T(R)dR},$$

$$= \frac{\int dR\Psi_T^*(R)\,\Psi_T(R)E_L(R)}{\int dR\Psi_T^*(R)\,\Psi_T(R)} \geq E_0, \quad (3.4)$$

$E_T$ is an energy expectation value for a system with Hamiltonian operator $\widehat{H}$; $\Psi_T(R) \equiv \Psi_T(R, a)$ is a trial wave function depending on all the particle coordinates possibly including spin, which are represented by the vector $R$, $a$ is a set of variational parameters, $E_L(R) = \Psi_T^{-1}(R)\widehat{H}\,\Psi_T(R)$ is the "local energy" of the trial wave function and $E_0$ is the exact ground-state energy. As the quality of the trial wave function improves, $E_T$ becomes closer to $E_0$.

The essence of VMC is the sampling of a distribution proportional to $|\Psi_T(R)|^2$ where $\Psi_T(R)$ is a given trial wave function, a function of the 3N-dimensional coordinates $R$. The expectation values of non differential operators may be sampled as [32]:

$$\langle \widehat{O} \rangle = \frac{\int \widehat{O}(R)|\Psi_T(R)|^2 d^{3N}R}{\int |\Psi_T(R)|^2 d^{3N}R} = \frac{1}{M}\sum_{i=1}^{M}\widehat{O}(R_i). \quad (3.5)$$

Differential operators are only slightly more difficult, since we can write:

$$\langle \widehat{O} \rangle_{VMC} = \frac{\langle \Psi_T|\widehat{O}|\Psi_T \rangle}{\langle \Psi_T|\Psi_T \rangle} = \frac{\int [\widehat{O}\,\Psi_T(R)/\,\Psi_T(R)]|\Psi_T(R)|^2 d^{3N}R}{\int |\Psi_T(R)|^2 d^{3N}R} = \frac{1}{M}\sum_{i=1}^{M}\widehat{O}\,\Psi_T(R_i)/\,\Psi_T(R_i). \quad (3.6)$$

## 3.2 The Varitional Wave Function

A suitably parameterized varitional wave function $\Psi_V$ which depends upon several varitional parameters $(\alpha_1, \alpha_2, ..., \alpha_n)$ is used to calculate an upper bound to the ground-state energy [26]:

$$E_V = \frac{\langle \Psi_V|H|\Psi_V\rangle}{\langle \Psi_V|\Psi_V\rangle} \geq E_0. \qquad (3.7)$$

The parameters in $\Psi_V$ can be varied to minimize $E_V$ and the lowest value is taken as the approximate ground-state energy, then the best $\Psi_V$ can be used to evaluate other operators. The variational wave function has the general form [31,33]:

$$\Psi_V = \left(S \prod_{i<j} F_{ij}\right)\emptyset, \qquad (3.8)$$

where $S$ is a symmetrizing operator (since $F_{ij}$ do not commute), $F_{ij}$ are pair correlation operators and $\emptyset = \emptyset(JMTM_T)$ is the initial uncorrelated state which is antisymmetrized product of single-particle spin-isospin states with no spatial dependence [26]. For the $^3$H and $^4$He nuclei $\emptyset$ is simply taken to be [31]

$$\emptyset(^3H) = A|p\uparrow n\downarrow n\uparrow\rangle. \qquad (3.9)$$

$$\emptyset(^4He) = A|p\downarrow p\uparrow n\downarrow n\uparrow\rangle. \qquad (3.10)$$

The antisymmetrizing operator $A$ in (3.9) and (3.10) takes the form:

$$A = (1 + e_{12} + e_{23} + e_{31} + e_{12}e_{23} + e_{23}e_{12}), \qquad (3.11)$$

where $e_{ij}$ are spin-isospin exchange operators

$$e_{ij} = -\frac{1}{4}[1 + \sigma_i \cdot \sigma_j + \tau_i \cdot \tau_j + (\sigma_i \cdot \sigma_j)(\tau_i \cdot \tau_j)]. \qquad (3.12)$$

The pair correlation operators are taken to be:

$$F_{ij} = \sum_{p=1,8} f^p(r_{ij})O_{ij}^p = f_{ij}^c(r_{ij})\left(1 + \sum_{p=2,8} u_{ij}^p(r_{ij})O_{ij}^p\right), \qquad (3.13)$$

where $f_{ij}^c$ are the central pair correlation functions.

The general form of $\Psi_V$ is [23]

$$\Psi_V = \left\{S \prod_{i<j} f_{ij}^c(r_{ij})\left[1 + \sum_{p=2,6}\left(\prod_{k\neq i,j} f_{ijk}^p\right)u_{ij}^p(r_{ij})O_{ij}^p\right]\right\}\emptyset, \qquad (3.14)$$

where $f_{ijk}^p$ are the three-body correlations which represent the effect of the other particles on the $u^p$,

$$f_{ijk}^p = 1 - t_1\left(\frac{r_{ij}}{R_{ijk}}\right)^{t_2} exp(-t_3 R_{ijk}). \qquad (3.15)$$

Also, $R_{ijk} = r_{ij} + r_{jk} + r_{ki}$, $t_1$ and $t_2$ are variational parameters.

## 4. The Urbana Model Three Nucleon Interactions

Explicit treatments of the pions and deltas degrees of freedom, or implicit treatment via an effective three-body force (3BF), are different ways to approach the same physics. In approach-1, the nucleus is assumed to be consisting of nucleons only. As nucleon-nucleon (N-N) forces one uses one of the standard two-body force (2BF) potentials (Reid Soft Core, Paris, Bonn, …), supplemented by a phenomenological or microscopically derived 3BF such as the Tucson-Melbourne (TM) force. This approach is taken by the Los Alamos, Urbana, Tohoku groups [34-36].

The second approach referred to as "3-body force" is based on an explicit treatment of the non-nucleonic degrees of freedom in the ground state wave function. It can be included microscopically by allowing in the wave function for pions, deltas, and pairwise interactions with these additional constituents. This approach is taken by the Hannover group [37].

The Urbana model (UVII) three nucleon interaction is written as a sum of long-range two pion exchange and intermediate-range repulsive terms [23]

$$V_{ijk} = V_{ijk}^{FM} + V_{ijk}^{R} = V_{ijk}^{2\pi} + V_{ijk}^{R}. \qquad (4.1)$$

$$V_{ijk}^{FM} = \sum_{cyc} -0.0333 \left( \{\tau_i \cdot \tau_j, \tau_j \cdot \tau_k\} \{x_{ij}, x_{ik}\} + \frac{1}{4} [\tau_i \cdot \tau_j, \tau_j \cdot \tau_k][x_{ij}, x_{ik}] \right),$$

$$x_{ij} = T(r_{ij})S_{ij} + \sigma_i \cdot \sigma_j Y(r_{ij}),$$

$$V_{ijk}^{R} = \sum_{cyc} 0.0038 \, T^2(r_{ij}) T^2(r_{jk}). \qquad (4.2)$$

The $T(r)$ and $Y(r)$ are radial functions associated with the tensor and Yukawa parts of the one pion-exchange interaction:

$$Y(r) = \frac{e^{-\mu r}}{\mu r}(1 - e^{-br^2}),$$

$$T(r) = \left[1 + \frac{3}{\mu r} + \frac{3}{(\mu r)^2}\right](1 - e^{-br^2}). \qquad (4.3)$$

Here the pion mass $\mu = 0.7 fm^{-1}$, and $b = 2 fm^{-2}$. The $V_{ijk}^{2\pi}$ is the familiar Fujita-Miyazawa two-pion exchange operator, and is attractive. The $V_{ijk}^{R}$ is repulsive, and its strength $U_0$ is 0.0038 in model VII instead of 0.003 in model $V$.

The Coulomb interaction is taken as [38]

$$V_c(r_{ij}) = \frac{e^2}{4r_{ij}}(1 + \tau_{3,i})(1 + \tau_{3,j}) \left[1 - \frac{1}{48}e^{-x}(48 + 33x + 9x^2 + x^3)\right], \qquad (4.4)$$

and $x = \sqrt{12} r_{ij}/R_{cp}$, where $R_{cp}$ is the rms charge radius of the proton.

## 5. Results and Conclusions

The ground-state wave function of each nucleus is expanded in series in terms of the basis functions of the TISM with even number of quanta of excitations $N$. Accordingly, each one of these basis functions is then expanded in terms of the two-particle total fractional parentage coefficients which are products of orbital and spin-isospin coefficients in order to calculate the two-particle operators in the Hamiltonian. The ground-state of triton has total angular momentum $J = \frac{1}{2}$, isotopic spin $T = \frac{1}{2}$ and even parity, i.e. $(J^\pi, T) = \left(\frac{1}{2}^+, \frac{1}{2}\right)$. In Table-1 we present the used TISM basis functions for the ground-state of triton corresponding to number of quanta of excitations $N = 0, 2, 4, 6, 8, 10, 12, 14, 16, 18$ and $20$. The basis functions which reproduce in the final calculations weights $\leq 10^{-6}$ are eliminated and then the resulting nuclear wave functions are renormalized to unity.

The energy eigenvalues which result from the diagonalization of the Hamiltonian matrices for triton showed two accepted values for the energy of the state $\left(\frac{1^+}{2}, \frac{1}{2}\right)$, the lowest one belongs to the ground-state, and hence the negative value of the binding energy, and the highest belongs to the first-excited state energy: $E^*$, which is not yet assigned experimentaly.

In Tables-2, and 3 we present the triton ground-state wave function ($\Psi$), after eliminating the bases with weights $\leq 10^{-6}$ and then renormalizing $\Psi$, binding energy (B.E.), in MeV, root mean-square radius ($R$), in fm, and first excited–state energy ($E^*$), in MeV, together with the corresponding experimental values as functions of the oscillator parameter $\hbar\omega$ for both of the two potentials, respectively. The improvement values arised from using the Skyrme III three-body interaction are also given in these tables.

In Table-4 we present the best values of the triton B.E., $R$ and $E^*$, for the two potentials together with the Skyrme III three-body interaction in the case of $N$ = 0, 2, 4, 6, 8, 10, 12, 14, 16, 18 and 20. The values of the oscillator parameter $\hbar\omega$, in MeV, which produce the best fit to the experimental data are also given in this table. Moreover, the corresponding experimental values are also given in the last row of this table.

It is seen from Table-4 that the inclusion of the three-body interaction improved the calculated ground-state energy of triton as expected.

The ground state of $^4$He has total angular momentum quantum number $J = 0$, isotopic spin $T = 0$ and even parity and so is the first–excited state. The Hamiltonian matrices for the ground state of $^4$He are constructed and diagonalized with respect to the oscillator parameter $\hbar\omega$, which is allowed to vary in a large range of values $8 \leq \hbar\omega \leq 28$ MeV in order to obtain the minimum energy eigenvalues. In Table-5 we present the TISM basis functions for the ground-state of $^4$He corresponding to number of quanta of excitations $N$ = 0, 2, 4, 6, 8, 10, 12, 14, 16, 18 and 20. The basis functions which reproduce in the final calculations weights $\leq 10^{-6}$ are eliminated and then the resulting nuclear wave functions are renormalized to unity.

Since the range of values of the oscillator parameter $\hbar\omega$ is large we present only in Table-6 the resulting ground-state nuclear wave functions of $^4$He for both of the two potentials, together with the Skyrme III three-body interaction, at the values of $\hbar\omega$ which gave the best fit between the calculated ground-state characteristics of $^4$He and the correspondoing experimental values.

Table-1 TISM-basis functions for the triton nucleus with $0 \leq N \leq 20$. The basis functions which reproduce in the final calculations weights $\leq 10^{-6}$ are eliminated.

| $\Psi_i$ $i$ | $N$ | $\{\rho\}$ | $(\nu)$ | $[f]$ | $L$ | $S$ |
|---|---|---|---|---|---|---|
| 1 | 0 | {0} | (0) | [3] | 0 | 1/2 |
| 2 | 2 | {2} | (0) | [3] | 0 | 1/2 |
| 3 | 2 | {2} | (2) | [21] | 0 | 1/2 |
| 4 | 2 | {2} | (2) | [21] | 2 | 3/2 |
| 5 | 2 | {11} | (0)* | [1³] | 1 | 1/2 |
| 6 | 4 | {4} | (0) | [3] | 0 | 1/2 |
| 7 | 4 | {4} | (2) | [21] | 0 | 1/2 |
| 8 | 4 | {4} | (2) | [21] | 2 | 3/2 |
| 9 | 4 | {31} | (2) | [21] | 1 | 1/2 |
| 10 | 6 | {6} | (0) | [3] | 0 | 1/2 |
| 11 | 6 | {6} | (2) | [21] | 0 | 1/2 |
| 12 | 8 | {8} | (0) | [3] | 0 | 1/2 |
| 13 | 10 | {10} | (0) | [3] | 0 | 1/2 |
| 14 | 12 | {12} | (0) | [3] | 0 | 1/2 |
| 15 | 14 | {14} | (0) | [3] | 0 | 1/2 |
| 16 | 16 | {16} | (0) | [3] | 0 | 1/2 |
| 17 | 18 | {18} | (0) | [3] | 0 | 1/2 |
| 18 | 20 | {20} | (0) | [3] | 0 | 1/2 |

Table-2 Ground-State Characteristic of triton with $0 \leq N \leq 20$ for Pot-I alone and for Pot-I together with the Skyrme III three-body interaction. The basis functions which reproduce in the final calculations weights $\leq 10^{-6}$ are eliminated and then the resulting nuclear wave functions are renormalized to unity.

| Charact. | | $\hbar\omega$ (in MeV) | | | | | | | Exp. |
|---|---|---|---|---|---|---|---|---|---|
| | | 11 | 12 | 13 | 14 | 15 | 16 | 17 | |
| $\Psi$ | $\Psi_1$ | 0.8835 | 0.8929 | 0.8936 | 0.8939 | 0.9019 | 0.9013 | 0.9085 | |
| | $\Psi_2$ | 0.1113 | 0.0617 | 0.1218 | 0.0682 | 0.1819 | 0.0522 | 0.0485 | |
| | $\Psi_3$ | -.3433 | -.3334 | -.3207 | -.3223 | -.2168 | -.3108 | -.2883 | |
| | $\Psi_4$ | 0.0877 | 0.0861 | 0.0711 | 0.0838 | 0.0780 | 0.0812 | 0.0741 | |
| | $\Psi_5$ | 0.0200 | 0.0199 | 0.0201 | 0.0195 | 0.0284 | 0.0191 | 0.0177 | |
| | $\Psi_6$ | 0.1964 | 0.1953 | 0.1822 | 0.1952 | 0.2400 | 0.1971 | 0.2040 | |
| | $\Psi_7$ | 0.1437 | 0.1397 | 0.1409 | 0.1353 | 0.1264 | 0.1310 | 0.1214 | |
| | $\Psi_8$ | 0.1181 | 0.1196 | 0.1105 | 0.1208 | 0.1243 | 0.1225 | 0.1261 | |
| | $\Psi_9$ | 0.0050 | 0.0052 | 0.0049 | 0.0053 | 0.0157 | 0.0055 | 0.0058 | |
| | $\Psi_{10}$ | 0.0130 | 0.0087 | 0.0243 | 0.0093 | 0.0215 | 0.0103 | 0.0127 | |
| | $\Psi_{11}$ | -.0312 | -.0302 | -.0300 | -.0303 | -.0311 | -.0310 | -.0300 | |
| | $\Psi_{12}$ | 0.0519 | 0.0514 | 0.0528 | 0.0516 | 0.0541 | 0.0526 | 0.0561 | |
| | $\Psi_{13}$ | 0.0352 | 0.0332 | 0.0407 | 0.0294 | 0.0199 | 0.0279 | 0.0210 | |
| | $\Psi_{14}$ | 0.0491 | 0.0453 | 0.0531 | 0.0421 | 0.0397 | 0.0398 | 0.0375 | |
| | $\Psi_{15}$ | 0.0377 | 0.0370 | 0.0400 | 0.0363 | 0.0363 | 0.0387 | 0.0368 | |
| | $\Psi_{16}$ | 0.0318 | 0.0412 | 0.0353 | 0.0411 | 0.0242 | 0.0411 | 0.0365 | |
| | $\Psi_{17}$ | 0.0289 | 0.0343 | 0.0257 | 0.0341 | 0.0355 | 0.0216 | 0.0320 | |
| | $\Psi_{18}$ | 0.0222 | 0.0252 | 0.0278 | 0.0213 | 0.0271 | 0.0210 | 0.0341 | |
| B.E | | 7.9102 | 8.1145 | 8.2366 | 8.2915 | 8.2588 | 8.1535 | 8.0552 | 8.48 |
| B.E Improved | | 8.2311 | 8.3589 | 8.4301 | 8.4795 | 8.4567 | 8.4182 | 8.3448 | |
| $R$ | | 2.1134 | 1.9728 | 1.8876 | 1.8062 | 1.8225 | 1.8744 | 1.9615 | 1.75 |
| $R$ Improved | | 1.9589 | 1.8766 | 1.7999 | 1.7512 | 1.7688 | 1.8031 | 1.8567 | |
| $E^*$ | | 8.4943 | 8.5834 | 8.6712 | 8.7431 | 8.8254 | 8.9347 | 9.9764 | --- |
| $E^*$ Improved | | 8.2631 | 8.3496 | 8.4345 | 8.5212 | 8.6121 | 8.7214 | 8.8945 | |

Table-3 Ground-State Characteristic of Triton with $0 \leq N \leq 20$ for Pot-II alone and for Pot-II together with the Skyrme III three-body interaction. The basis functions which reproduce in the final calculations weights $\leq 10^{-6}$ are eliminated and then the resulting nuclear wave functions are renormalized to unity.

| Charact. | | $\hbar\omega$ (in MeV) | | | | | | | Exp. |
|---|---|---|---|---|---|---|---|---|---|
| | | 11 | 12 | 13 | 14 | 15 | 16 | 17 | |
| $\Psi$ | $\Psi_1$ | 0.8599 | 0.8795 | 0.8893 | 0.8906 | 0.8982 | 0.9029 | 0.9035 | |
| | $\Psi_2$ | 0.1961 | 0.1125 | 0.0960 | 0.0691 | 0.0395 | 0.0251 | 0.0498 | |
| | $\Psi_3$ | -.3507 | -.3415 | -.3222 | -.3222 | -.3126 | -.3026 | -.2919 | |
| | $\Psi_4$ | 0.1334 | 0.1365 | 0.1373 | 0.1370 | 0.1360 | 0.1340 | 0.1309 | |
| | $\Psi_5$ | 0.0574 | 0.0596 | 0.0614 | 0.0626 | 0.0633 | 0.0635 | 0.0632 | |
| | $\Psi_6$ | 0.1812 | 0.1773 | 0.1750 | 0.1735 | 0.1740 | 0.1760 | 0.1791 | |
| | $\Psi_7$ | 0.1392 | 0.1357 | 0.1321 | 0.1283 | 0.1245 | 0.1204 | 0.1160 | |
| | $\Psi_8$ | 0.1266 | 0.1271 | 0.1277 | 0.1282 | 0.1291 | 0.1300 | 0.1309 | |
| | $\Psi_9$ | 0.0241 | 0.0250 | 0.0258 | 0.0262 | 0.0267 | 0.0269 | 0.0269 | |
| | $\Psi_{10}$ | 0.0207 | 0.0192 | 0.0210 | 0.0161 | 0.0200 | 0.0302 | 0.0403 | |
| | $\Psi_{11}$ | -.0284 | -.0265 | -.0247 | -.0228 | -.0213 | -.0197 | -.0181 | |
| | $\Psi_{12}$ | 0.0472 | 0.0456 | 0.0447 | 0.0563 | 0.0449 | 0.0460 | 0.0477 | |
| | $\Psi_{13}$ | 0.0403 | 0.0354 | 0.0313 | 0.0410 | 0.0237 | 0.0203 | 0.0169 | |
| | $\Psi_{14}$ | 0.0388 | 0.0312 | 0.0299 | 0.0334 | 0.0285 | 0.0308 | 0.0288 | |
| | $\Psi_{15}$ | 0.0354 | 0.0247 | 0.0310 | 0.0303 | 0.0305 | 0.0253 | 0.0298 | |
| | $\Psi_{16}$ | 0.0324 | 0.0371 | 0.0364 | 0.0486 | 0.0371 | 0.0331 | 0.0377 | |
| | $\Psi_{17}$ | 0.0228 | 0.0263 | 0.0298 | 0.0325 | 0.0329 | 0.0261 | 0.0341 | |
| | $\Psi_{18}$ | 0.0213 | 0.0221 | 0.0214 | 0.0246 | 0.0216 | 0.0210 | 0.0112 | |
| B.E. | | 7.8180 | 8.0137 | 8.1393 | 8.2173 | 8.2043 | 8.1863 | 8.1182 | 8.48 |
| B.E. Improved | | 8.0213 | 8.2159 | 8.3510 | 8.4311 | 8.4176 | 8.3992 | 8.3293 | |
| $R$ | | 2.2209 | 2.1303 | 2.0520 | 1.9240 | 1.9799 | 2.0265 | 2.1088 | 1.75 |
| $R$ Improved | | 2.0464 | 1.9654 | 1.8909 | 1.7729 | 1.8244 | 1.8674 | 1.9432 | |
| $E^*$ | | 8.5323 | 8.6042 | 8.6897 | 8.7754 | 8.8779 | 8.9843 | 9.1565 | --- |
| $E^*$ Improved | | 8.4645 | 8.5272 | 8.5923 | 8.6654 | 8.7523 | 8.8413 | 9.0623 | |

Table-4 The best values of the ground state characteristics of triton by using the two potentials together with the Skyrme III three-body interaction. The corresponding experimental values and the values of the oscillator parameter $\hbar\omega$ are also given.

| Characteristic Case | B.E. (MeV) | $E^*$ (MeV) | $R$ (fm) | $\hbar\omega$ (MeV) |
|---|---|---|---|---|
| Pot-I +Skyrme III Pot | 8.4795 | 8.5212 | 1.7512 | 14.0 |
| Pot-II + Skyrme III Pot | 8.4311 | 8.6654 | 1.7729 | 14.0 |
| Exp. | 8.48 [39] | --- | 1.75 [39] | --- |

Table-5 The TISM bases of the ground-state of $^4$He. The basis functions which reproduce in the final calculations weights $\leq 10^{-6}$ are eliminated.

| $\Psi_i$ $i$ | N | {ρ} | (ν) | [f] | L | S |
|---|---|---|---|---|---|---|
| 1 | 0 | {0} | (0) | [4] | 0 | 0 |
| 2 | 2 | {2} | (0) | [4] | 0 | 0 |
| 3 | 2 | {2} | (0) | [22] | 0 | 0 |
| 4 | 2 | {2} | (2) | [22] | 2 | 2 |
| 5 | 2 | {11} | (1) | [211] | 1 | 1 |
| 6 | 4 | {4} | (0) | [4] | 0 | 0 |
| 7 | 4 | {4} | (2) | [22] | 0 | 0 |
| 8 | 4 | {4} | (2) | [22] | 2 | 2 |
| 9 | 4 | {31} | (2) | [31] | 1 | 1 |
| 10 | 6 | {6} | (0) | [4] | 0 | 0 |
| 11 | 6 | {6} | (2) | [22] | 2 | 2 |
| 12 | 8 | {8} | (0) | [4] | 0 | 0 |
| 13 | 8 | {8} | (2) | [22] | 2 | 2 |
| 14 | 10 | {10} | (0) | [4] | 0 | 0 |
| 15 | 10 | {10} | (2) | [22] | 2 | 2 |
| 16 | 12 | {12} | (0) | [4] | 0 | 0 |
| 17 | 14 | {14} | (0) | [4] | 0 | 0 |
| 18 | 16 | {16} | (0) | [4] | 0 | 0 |
| 19 | 18 | {18} | (0) | [4] | 0 | 0 |
| 20 | 20 | {20} | (0) | [4] | 0 | 0 |

Table-6 Ground-state nuclear wave functions of $^4$He for the two potentials together with the Skyrme III three-body interaction. The basis functions which reproduce in the final calculations weights $\leq 10^{-6}$ are eliminated and then the resulting nuclear wave functions are renormalized to unity.

| $\Psi_i$ $i$ | Pot-I + Skyrme III potential | Pot-II + Skyrme III potential |
|---|---|---|
| 1 | 0.9602 | 0.8764 |
| 2 | -0.0616 | 0.1528 |
| 3 | -0.0137 | -0.0671 |
| 4 | 0.1521 | 0.1723 |
| 5 | 0.0521 | -0.1129 |
| 6 | 0.1204 | 0.1784 |
| 7 | 0.0926 | 0.1223 |
| 8 | -0.0044 | -0.0920 |
| 9 | 0.0232 | 0.1540 |
| 10 | 0.0019 | 0.0930 |
| 11 | 0.0810 | -0.0997 |
| 12 | 0.0211 | 0.1305 |
| 13 | -0.0932 | 0.0971 |
| 14 | 0.0665 | -0.1453 |
| 15 | -0.0298 | -0.1001 |
| 16 | 0.0107 | 0.0453 |
| 17 | 0.0167 | 0.0184 |
| 18 | 0.0340 | 0.0443 |
| 19 | 0.0286 | 0.0370 |
| 20 | 0.0324 | 0.0282 |
| $\hbar\omega$ (MeV) | 17 | 17 |

The energy eigenvalues which result from the diagonalization of the Hamiltonian matrices for $^4$He showed two accepted values for the energy of the state $(0^+, 0)$, the lowest one belongs to the ground-state and the highest belongs to the first-excited state: $E^*$. The obtained nuclear wave functions are used to calculate the root mean-square radius of $^4$He. We present in Table-7 the best values of the $^4$He binding energy (B.E.), in MeV, root mean-square radius ($R$), in fm, and first-excited state energy eigenvalue ($E^*$), in MeV, for the two potentials alone and the two potentials together with the Skyrme III three-body interaction, in the case of $N = 0, 2, 4, 6, 8, 10, 12, 14, 16, 18$ and $20$. The corresponding experimental values and previous results by using the Gogny, Pires and De Tourreil (GPT)-potential [19] with the TISM bases corresponding to $0 \leq N \leq 10$ [10] are also given in this table. In Table-7 we present also previous results obtained by using large-basis shell model with a multivalued G-matrix effective interaction [24]. The values of the oscillator parameter $\hbar\omega$, which reproduced the minimum energy eigenvalues in each case, are also given in Table-7.

Table-7 The best values of the ground state characteristics of $^4$He by using the two potentials alone and the two potentials together with the Skyrme III three-body interaction. The corresponding experimental values and the values of the oscillator parameter $\hbar\omega$ are also given. Previous results by using the GPT-potential and the G-matrix method are also given.

| Characteristic<br>Case | B. E.<br>(MeV) | $R$<br>(fm) | $E^*$<br>(MeV) | $\hbar\omega$<br>(MeV) |
|---|---|---|---|---|
| Pot-I | 27.895 | 1.534 | 20.654 | 17 |
| Pot-I + Skyrme III-Pot | 28.286 | 1.468 | 20.354 | |
| Pot-II | 27.554 | 1.687 | 21.515 | 17 |
| Pot-II + Skyrme III-Pot | 28.0735 | 1.512 | 21.222 | |
| GPT-Potl, $N \leq 10$ [10] | 25.480 | 1.70 | 20.70 | 17 |
| G-matrix [24] | 26.459 | 1.492 | 21.82 | 14 |
| Experiment | 28.3 [40] | 1.46 [27] | 20.21 [26] | _ |

It is seen from Table-7 that the calculated values of the binding energy of $^4$He are in excellent agreement with the corresponding experimental value for the two potentials. Concerning the calculated values of the root mean-square radius and the first excited-state energy of $^4$He the first potential gave results in better agreement with the corresponding experimental values rather than the second potential.

Least-square fit to the obtained results of the binding energy, the root mean-suare radius and the first excited-state energy of $^3$H and $^4$He by using Pot I (Pot II) alone and the two potentials together with the three-body potential, are carried out as functions of the second degree in the value of the oscillator parameter $\hbar\omega$, and the the results are

$$\text{characteristic value} = a_0 + a_1[\hbar\omega] + a_2([\hbar\omega])^2, \tag{5.1}$$

where $[\hbar\omega]$ is the value of $\hbar\omega$. The values of the parameters $a_0$, $a_1$ and $a_2$, for $^3\text{H}$ and $^4\text{He}$, are given in Tables-8 and -9, respectively.

Table-8 Least-square fit to the obtained results for $^3\text{H}$

| Parameter<br>Characteristic | $a_0$ | $a_1$ | $a_2$ |
|---|---|---|---|
| B.E<br>Pot-I | −0.000157 MeV | 1.166 MeV | −0.0409 MeV |
| B.E<br>Pot-I + Skyrme III-Pot | 0.787 MeV | 1.0815 MeV | −0.038 MeV |
| R<br>Pot-I | 6.9 fm | −0.71 fm | 0.0247 fm |
| R<br>Pot-I + Skyrme III-Pot | 5.19 fm | −0.475 fm | 0.0164 fm |
| $E^*$<br>Pot-I | 8.35 MeV | −0.037 MeV | 0.0046 MeV |
| $E^*$<br>Pot-I + Skyrme III-Pot | 9.2 MeV | −0.203 MeV | 0.011 MeV |
| B.E<br>Pot-II | 2.41 MeV | 0.82 MeV | −0.029 MeV |
| B.E<br>Pot-II + Skyrme III-Pot | 2.43 MeV | 0.81 MeV | −0.0271 MeV |
| R<br>Pot-II | 6.61 fm | −0.657 fm | 0.023 fm |
| R<br>Pot-II + Skyrme III-Pot | 6.05 fm | −0.6 fm | 0.021 fm |
| $E^*$<br>Pot-II | 8.85 MeV | −0.112 MeV | 0.007 MeV |
| $E^*$<br>Pot-II + Skyrme III-Pot | 9.55 MeV | −0.22 MeV | 0.011 MeV |

Table-9 Least-square fit to the obtained results for $^4\text{He}$

| Parameter<br>Characteristic | $a_0$ | $a_1$ | $a_2$ |
|---|---|---|---|
| B.E<br>Pot-I | 4.423 MeV | 2.73 MeV | −0.08 MeV |
| B.E<br>Pot-I + Skyrme III-Pot | 8.44 MeV | 2.358 MeV | −0.07 MeV |
| R<br>Pot-I | 4.67 fm | −0.387 fm | 0.0117 fm |
| R<br>Pot-I + Skyrme III-Pot | 4.6 fm | −0.38 fm | 0.0115 fm |
| $E^*$<br>Pot-I | 44.3 MeV | −2.78 MeV | 0.081 MeV |
| $E^*$ Pot-I + Skyrme III-Pot | 46.1 MeV | −3.05 MeV | 0.09 MeV |
| B.E Pot-II | 3.7 MeV | 2.77 MeV | −0.08 MeV |
| B.E Pot-II + Skyrme III-Pot | 4.13 MeV | 2.81 MeV | −0.082 MeV |
| R Pot-II | 5.32 fm | −0.45 fm | 0.0136 fm |
| R Pot-II + Skyrme III-Pot | 5.37 fm | −0.47 fm | 0.0142 fm |
| $E^*$ Pot-II | 44.65 MeV | −2.674 MeV | 0.0795 MeV |
| $E^*$ Pot-II + Skyrme III-Pot | 42.43 MeV | −2.47 MeV | 0.072 MeV |

According to the above results, the variations of the binding energy (B.E.), root mean-square radius ($R$), and first excited–state energy ($E^*$), with respect to the oscillator parameter $\hbar\omega$ (i.e. the dependence of the obtained results on the used model and its wave function) for each potential, with and without improvements arised from using the Skyrme III three-body interaction, are given in Figs. 1-12, for the nuclei $^3$H and $^4$He. The figures show minima for the values of the root mean-square radius and the ground-stae energy (the negative value of the binding energy) for the the triton and the helium nuclei, in agreement with the basic property of the used model. The figures gave no physical inerpretation for the first excited-state energy of triton, but for the helium nucleus minima have been occurred for the two cases of the potentials. This is due to the fact that the $^4$He-excited state $(0^+, 0)_2$ is observed experimentally. Moreover, it is seen from Table-2 and Figs.1, 2 and 3, for $^3$H, that the differences arising from using potential-I alone and by adding to it the Skyrme III three-body interaction in the cacluations of the binding energy, the root mean-square radius and the first excited–state energy are 0.188 MeV, -0.055 fm and 0.222 MeV, respectively. We see also that at $\hbar\omega = 14$ MeV the improvement calculations arised from using the Skyrme III three-body interaction together with Pot-I gave values close to the experimintal ones, where the differences between them are 0.0005 MeV in the binding energy and -0.0012 fm in the root mean-square radius of $^3$H. Furthermore, it is seen from Table-3 and Figs.4, 5 and 6 that the differences arising from using potential-II alone and by adding to it the Skyrme III three-body interaction in the cacluations of the binding energy, the root mean-square radius and the first excited–state energy of $^3$H are 0.2138 MeV, -0.1511 fm and 0.11 MeV, respectively. We can see also that at $\hbar\omega = 14$ MeV the improvement calculations arised from using the Skyrme III three-body interaction together with Pot-II gave values close to the experimintal ones, where the differences between them are 0.0489 MeV in the binding energy of $^3$H and -0.1229 fm in the value of the root mean-square radius of $^3$H.

Also, for $^4$He it is seen from Table-7 and Figs. 7, 8 and 9 that the differences between using potential-I alone and by adding to it the Skyrme III three-body interaction in the binding energy, root mean-square radius and first excited–state energy are 0.391 MeV, -0.066 fm and -0.3 MeV, respectively. We see also that at $\hbar\omega = 17$ MeV, the improvement calculations arised from using the Skyrme III three-body interaction together with Pot-I gave values close to the experimintal ones where the differences between them are 0.014 MeV in the binding energy of $^4$He, -0.008 fm in the root mean-square radius of $^4$He and - 0.0144 MeV in the first excited–state energy. Also, it is seen from Table-7 and Figs. 10, 11 and and 12, for $^4$He, that the differences between using potential-II alone and by adding to it the Skyrme III three-body interaction in the binding energy, the root mean-square radius and the first excited–state energy are 0.5195 MeV, -0.175 fm and -0.293 MeV, respectively. We see also that at $\hbar\omega = 17$ MeV the improvement calculations arised from using the Skyrme III three-body interaction together with Pot-II gave values, for $^4$He, close to the experminal ones where the differences between them are 0.2265 MeV in the binding energy, -0.052 fm in the root mean-square radius and -1.012 MeV in the first excited–state energy.

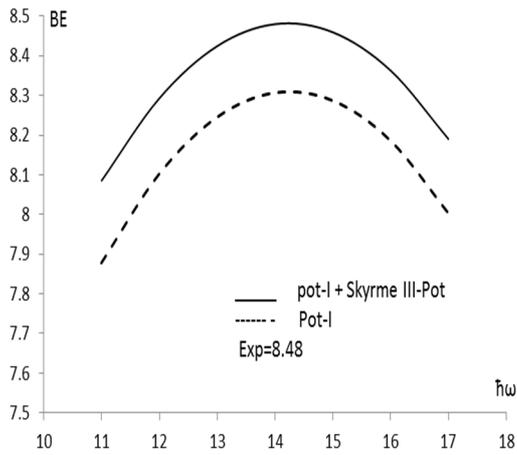

Fig. 1

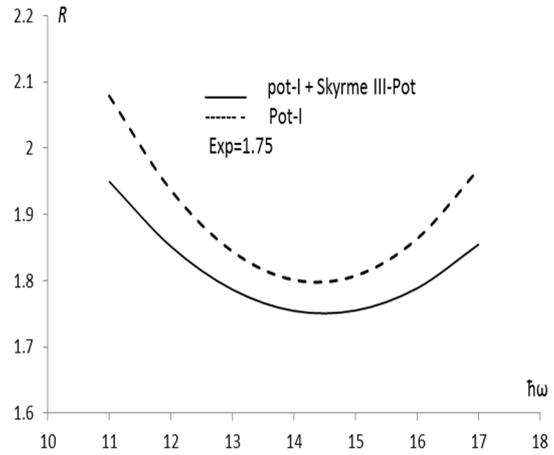

Fig. 2

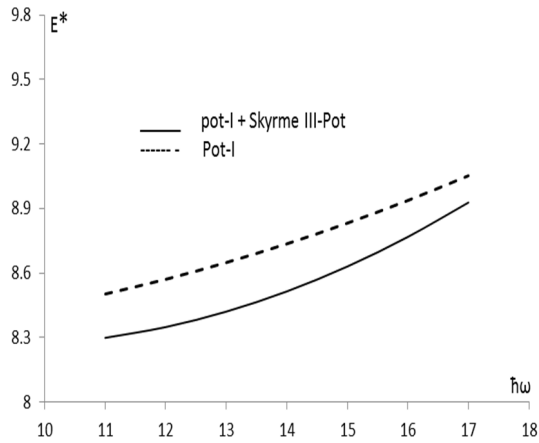

Fig. 3

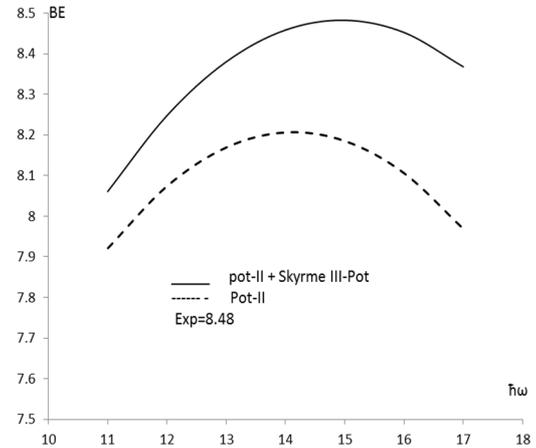

Fig. 4

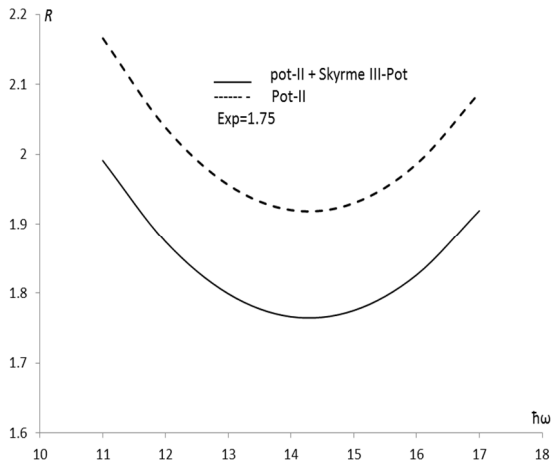
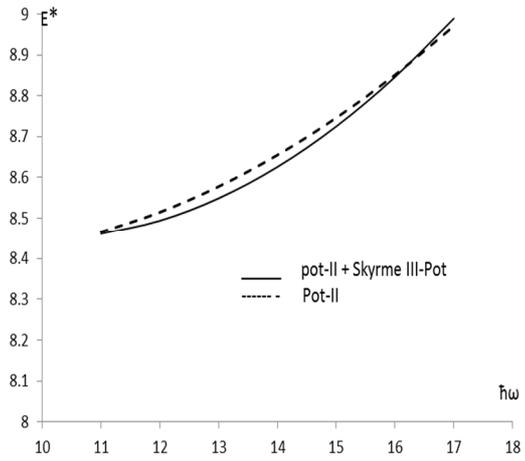

Fig. 5                                                  Fig. 6

Figs.1, 2, 3, 4, 5, and 6 Variations of the $^3$H binding energy (B.E.), root mean-square radius ($R$), and first excited–state energy ($E^*$), with the oscillator parameter $\hbar\omega$ for each potential (Pot-I and Pot-II), respectively. The improved results arising from using the Skyrme III three-body interaction are also given.

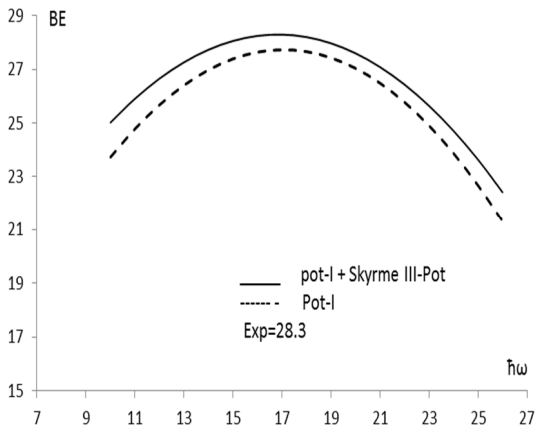
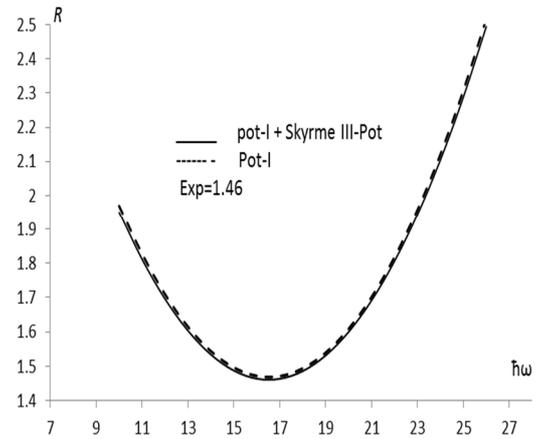

Fig. 7                                                  Fig. 8

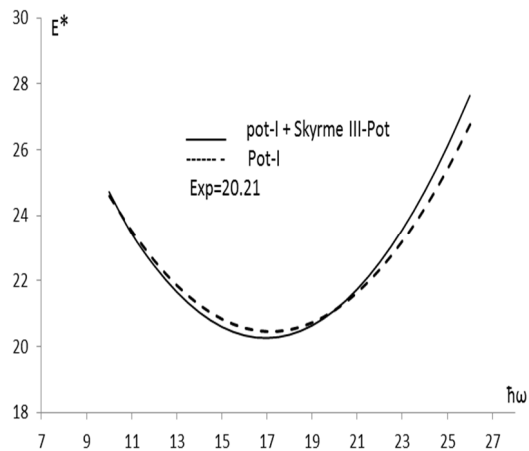
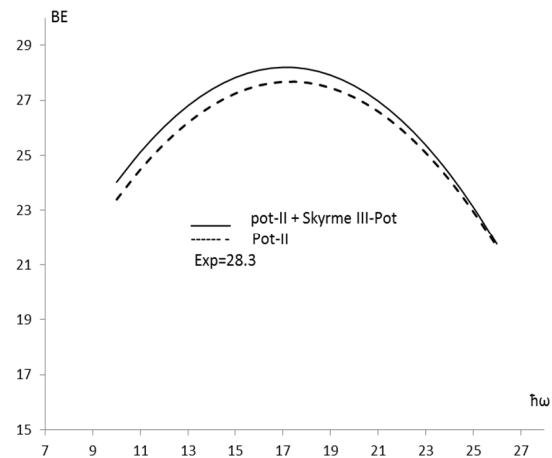

Fig. 9                                                  Fig. 10

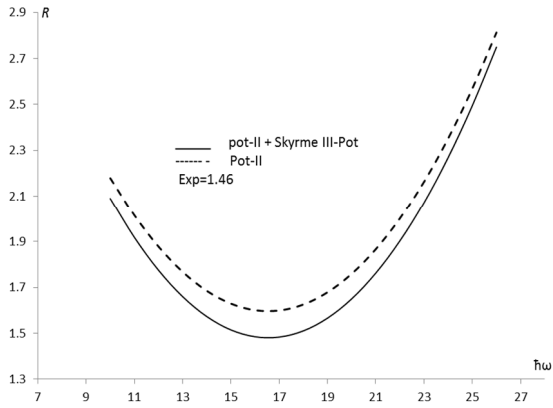
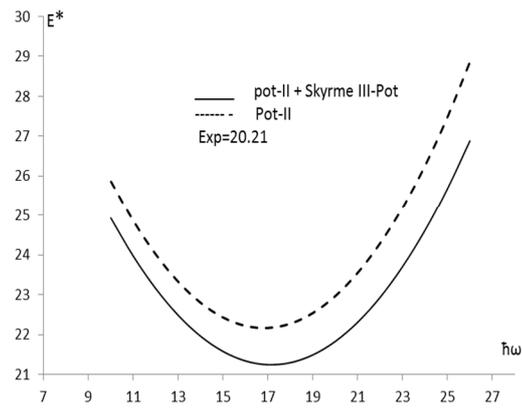

Fig. 11                                   Fig. 12

Figs. 7, 8, 9, 10, 11, and 12 Variations of the binding energy (B.E.), root mean-square radius ($R$), and first excited–state energy ($E^*$), of $^4$He with the oscillator parameter $\hbar\omega$ for each potential with the improvement values arised from using the Skyrme III three-body interaction.

In the Monte Carlo calculations the role of trial wave function is very important in order to obtain good results for the binding energy and the root mean-square radius of the considered nucleus as well as to accelerate the convergence of the resulting integrals. The using computer program enables us to make good search for such trial wave function. In Table-10 we present the triton variational parameters for the VMC method.

The optimized variational parameters that we have obtained for the triton with the Reid $V_8$ interaction are listed in Table-10. Once the optimum wave function has been determined, a set of Monte Carlo calculations should be undertaken to determine all of the expectation values. The results obtained with this wave function are summarized in Table-11. Table-11 presents all of the energy expectation values, as well as point-particle root mean square radii of the neutron and proton density. In this Table, $V_{ij}$ gives the total nucleon-nucleon potential energy, and $T_i$ is the total kinetic energy. The two-body potential is also split into the $V_6$ contribution and the remaining L.S terms ($V_b$)

In addition, the contribution of the various pieces of the Urbana model (UVII) three-nucleon interaction (TNI) are given in this Table. The commutator and anti-commutator pieces of the two-pion-exchange TNI are listed as $V3_c$ and $V3_a$ respectively, and the short-ranged repulsive piece is $V3_u$. Each of these terms is a very small fraction of the total potential energy, yet they constitute a significant part of the binding energy of the triton nucleus.

The contribution of the Urbana model (UVII) three-nucleon interaction (TNI) to the expectation value of the binding energy of triton is 0.76 MeV which represents 10% of the binding energy of triton when we use the Reid $V_8$ two-body interaction only. This increase depends on the exact 2BF and 3BF used. The inclusion of the 3BF improves the Coulomb energy difference $^3$H-$^3$He. The 3BF leads to a change of the proton point density [41]. This change of the point density due to the 3BF is very desirable one.

Table-10 Triton variational parameters for the VMC method

|  | $^1S_0$ | $^1P_1$ | $^3S_1$ | $^3P_1$ | $^3D_1$ | $^3F_J$ |
|---|---|---|---|---|---|---|
| $E_{S,T}$ | 6.00 | 2.00 | 12.00 | 6.00 | --- | --- |
| $\eta_T$ |  |  |  |  | 0.026 | 010.- |
| $C_x$ | 1.00 | 1.00 | 3.00 | 3.00 | 2.00 | 2.00 |
| $a_x$ | 0.40 | 0.40 | 0.40 | 0.40 | 0.40 | 0.40 |
| $R_x$ | 1.00 | 1.00 | 2.80 | 2.80 | 3.60 | 3.60 |
| $\alpha_x$ | 1.00 | 0.92 | 0.92 | 0.92 | 0.92 | 0.92 |
|  | $t1_{ST}$ | $t1_{tT}$ | $t2_{ST}$ | $t2_{tT}$ | $t3_{ST}$ | $t3_{tT}$ |
|  | 10.00 | 10.00 | 4.00 | 4.00 | 0.05 | 0.05 |

Table-11 The calculated ground-state energy of triton by using the variational Monte Carlo method with two- and three-body interactions.

| Characteristics | Expectation Value | Statistical Error |
|---|---|---|
| $T_i + V_{ij}$ | -7.59 | 0.03 |
| $T_i + V_{ij} + V_{ijk}$ | -8.35 | 0.06 |
| $T_i$ | 51.15 | 0.53 |
| $V_{ij}$ | -58.75 | 0.53 |
| $V_6$ | 59.43 | 0.53 |
| $V_b$ | .068 | 0.07 |
| $V_{ijk}$ | -0.75 | 0.04 |
| $V3_a$ | -0.57 | 0.03 |
| $V3_c$ | -0.39 | 0.01 |
| $V3_c$ | 0.20 | 0.006 |
| $\langle r_i^2 \rangle$ proton | 1.57 | 0.001 |
| $\langle r_i^2 \rangle$ neutron | 1.64 | 0.001 |

In Table-12 we present the $^4$He variational parameters. The calculated ground-state energy of helium by using three-body interaction, together with the two-body interaction are given in Table-13.

Table-12 $^4$He variational parameters for the VMC method

|  | $^1S_0$ | $^1P_1$ | $^3S_1$ | $^3P_1$ | $^3D_1$ | $^3F_J$ |
|---|---|---|---|---|---|---|
| $E_{S,T}$ | 6.00 | 2.00 | 12.0 | 6.00 | --- | --- |
| $\eta_T$ |  |  |  |  | .026 | -.010 |
| $C_x$ | 1.00 | 1.00 | 3.00 | 3.00 | 2.00 | 2.00 |
| $a_x$ | 0.40 | 0.40 | 0.40 | 0.40 | 0.40 | 0.40 |
| $R_x$ | 1.00 | 1.00 | 2.80 | 2.80 | 3.60 | 3.60 |
| $\alpha_x$ | 1.00 | 0.80 | 0.80 | 0.80 | 0.80 | 0.80 |
|  | $t1_{ST}$ | $t1_{tT}$ | $t2_{ST}$ | $t2_{tT}$ | $t3_{ST}$ | $t3_{tT}$ |
|  | 10.0 | 10.0 | 4.00 | 4.00 | 0.05 | 0.05 |

Table-13 The calculated ground-state energy of helium $^4$He by using the variational Monte Carlo method with two- and three-body interactions.

| Characteristics | Expectation Value | Statistical Error |
|---|---|---|
| $T_i + V_{ij}$ | -25.30 | 0.05 |
| $T_i + V_{ij} + V_{ijk}$ | -28.296 | 0.07 |
| $T_i$ | 75.025 | 0.39 |
| $V_{ij}$ | -100.32 | 0.42 |
| $V_6$ | -102.37 | 0.39 |
| $V_b$ | 1.43 | 0.04 |
| $V_{ijk}$ | -2.99 | 0.03 |
| $V3_a$ | -2.65 | 0.03 |
| $V3_c$ | -1.74 | 0.01 |
| $V3_c$ | 1.41 | 0.018 |
| $\langle r_i^2 \rangle$ proton | 1.92 | 0.001 |
| $\langle r_i^2 \rangle$ neutron | 1.92 | 0.001 |

Similar conclusions can be given for the nucleus $^4$He. The contribution of the Urbana model (UVII) three-nucleon interaction (TNI) to the expectation value of the binding energy of helium is 2.996 MeV which represents 9% of the binding energy of helium when we use the modified Reid $V_8$ two-body interaction only. It is seen from the second row of Table-13 that the inclusion of the three-nucleon interaction improved the calculated ground-state energy of helium as expected.

**References**
[1] I. Talmi, **Simple Models of Complex Nuclei**, Harwood Academic publishers (1993).
[2] V. V. Vanagas, **Algebraic Methods in Nuclear Theory**, Mintis. Vilnius (1971).
[3] V. Bargman and M. Moshinsky, Nucl. Phys., **18**: 697 (1960); **23**: 177 (1961),S. P. Kramer and Moshinsky, Nucl. Phys., **82**: 241 (1966).
[4] M. Kretzschmar, Z. Phys. **157**: 433 (1960); **158**: 284 (1960).
[5] S. B. Doma, Bull. Acad. Sci. Georgian SSR, **74**: 585 (1974).
[6] S. B. Doma, T. I. Kopaleyshvili and I. Z. Machabeli, Yadernaya Physica (Sov. J. Nucl. Phys.) **21**: 720 (1975).
[7] S. B. Doma and A. M. El-Zebidy, International Journal of Modern Physics E, **14**(2): 189 (2005).
[8]S. B. Doma, A. M. El-Zebidy and M. A. Abdel-Khalik, J. Phys. G: Nucl. Part. Phys. **34**: 27 (2007).
[9] S. B. Doma, Indian J. Pure Appl. Math., **12**(6): 769 (1981); **12**(12): 468 (1981).
[10] S. B. Doma, Helvetica Physica Acta, **58**: 1072 (1985).
[11] V. R. Pandharipande, Nucl. Phys. **A446**: 189c (1985).
[12] M. Viviani, L. Girlanda, A. Kievsky, and L. E. Marcucci, Phys. Rev. Lett. **111**, 172302 (2013).
[13] D. Lonardoni, S. Gandolfi, and F. Pederiva, Phys. Rev. C **87**, 041303(R) (2013).
[14] D. Lonardoni, F. Pederiva, and S. Gandolfi, Phys. Rev. C **89**, 014314 (2014).
[15] A. Cipollone, C. Barbieri, and P. Navrátil, Phys. Rev. Lett. **111**, 062501 (2013).
[16] R. B. Wiringa, R. Schiavilla, Steven C. Pieper, and J. Carlson, Phys. Rev. C **89**, 024305 (2014).

[17] Bruce R. Barrett, Petr Navrátil and James P. Vary, Progress in Particle and Nuclear Physics, **69**: 131 (2013).
[18] Christian Forssén, Petr Navrátil and Sofia Quaglioni, **49**: 11 (2011).
[19] D. Gogny, P. Pires and R. De Tourreil, Phys. Lett., **32B**: 591 (1970).
[20] S. B. Doma, N. A. El-Nohy and K. K. Gharib, Helvetica Physica Acta,**69**: 90 (1996).
[21] M. Beiner, H. Flocard, N. Van Giai et al. Nucl. Phys. A238 (1975).
[22] Reid R. V., Ann. Phys., **50**, 411 [1968].
[23] R. Schiavilla, V. R. Pandharipande, and R. B. Wiringa, Nucl. Phys. **A449**: 219 (1986).
[24] S. B. Doma, High Energy Physics and Nuclear Physics, **26**(9): 941 (2002).
[25] S. B. Doma, International Journal of Modern Physics E, **12**(3): 421 (2003).
[26] R. B. Wiringa, Phys. Rev. C**43**, No. 4: 1585 (1991).
[27] S. B. Doma and F. N. El-Gammal, Journal Applied Mathematics & Information Sciences, **5**(3): 315S (2011).
[28] S. B. Doma and F. N. El-Gammal, Acta Physica Polonica A, **122** (1): 42 (2012).
[29] S. B. Doma and F. N. El-Gammal, Journal of Theoretical and Applied Physics, **6**: 28 (2012).
[30] R. B. Wiringa and V. R. Pandharipande, Nucl. Phys. **A317**: 1 (1979).
[31] J. A. Carlson and R. B. Wiringa, **Computational Nuclear Physics1**: 171, Springer-Verlag, Berlin (1991).
[32] J. B. Shumway, JR. **Quantum Monte Carlo Simulations of Electrons and Holes**, Ph. D. Thesis, Univ. of Illinois (1999).
[33] I. E. Lagaris and V. R. Pandharipande, Nucl. Phys. **A359**: 349 (1981).
[34] R. B. Wiringa, J. L. Friar, B. F. Gibson, G. L. Payne and C. R. Chen, Phys. Lett. **143B**: 273 (1984)
[35] C. R. Chen, G. L. Payne, J. L. Friar and B. F. Gibson, preprint LA-UR, 85: 1472
[36] S. Ishikawa, T. Sasakwa, T. Sawada and T. Ueda, Phys. Rev. Lett. **53**: 1877 (1984).
[37] P. U. Sauer, Prog. Part. Nucl. Phys. **16**: 35 (1986).
[38] N. Auerbach et al., Rev. Mod. Phys. **44**: 48 (1972).
[39]D. C. Zheng, J. P. Vary and B. R. Barrett, Physical Review C, 50: 2841 (1994).
[40] Samuel S. M. Wong, **Introductory Nuclear Physics**, Wiley-Vch Verlag Gmbl **I** & Co. KGaA, Weinheim (2004).
[41] I. Sick, Proc. of the International Symposium, George Washington University, Washington D. C., April 24-26 (1986).